\documentclass[11pt]{article}
\usepackage{amssymb}
\usepackage{euscript}
\def\l{\lambda}
\def\a{{\alpha}}
\def\b{{\beta}}
\def\d{\partial}
\def\l{\lambda}
\def\x{{\rm x}}

\def\bra#1{\langle #1 |}
\def\ket#1{|#1 \rangle}
\def\0{\nonumber}

\def\Det{{\rm Det}}

\def\tr{{\rm tr}}

\newcommand\ES{\EuScript{S}}
\newcommand\EV{\EuScript{V}}

\newcommand\T{\EuScript{T}}

\newcommand\X{\EuScript{X}}

\newcommand\N{{\cal{N}}}

\newcommand\I{\mathbb{I}}
\newcommand\ee{\end{eqnarray}}      
\newcommand\be{\begin{eqnarray}}
\newcommand\ba{\begin{array}}           
\newcommand\ea{\end{array}}
\newcommand\eeq{\end{equation}}     
\newcommand\beq{\begin{equation}}

\begin{document}
\begin{flushright}
SISSA/4/06/EP\\ULB--TH/06-03\\ hep-th/0602015
\end{flushright}

\vspace{.1in}
\begin{center}
{\LARGE\bf Bubbling AdS and Vacuum String Field Theory}
\end{center}
\vspace{0.1in}
\begin{center}
L. Bonora $^{(a)}$\footnote{ bonora@sissa.it}, C.Maccaferri
$^{(b)}$ \footnote{carlo.maccaferri@ulb.ac.be}, R.J.Scherer Santos
$^{(a)}$\footnote{scherer@sissa.it}, D.D.Tolla $^{(a)}$
\footnote{tolla@sissa.it}\\
\vspace{7mm}

$^{(a)}$ {\it International School for Advanced Studies (SISSA/ISAS)\\
Via Beirut 2--4, 34014 Trieste, Italy, and INFN, Sezione di
Trieste} \vspace{5mm}

$^{(b)}$  {\it Physique Th\'eorique et Math\'ematique,
Universit\'e Libre de Bruxelles \& International Solvay
Institutes, ULB Campus Plaine C.P. 231, B--1050 Bruxelles,
Belgium}
\end{center}
\vspace{0.1in}
\begin{center}
{\bf Abstract}
\end{center}
We show that a family of 1/2--BPS states of $\N=4$ SYM is in
correspondence with a family of classical solutions
of VSFT with a $B$--field playing the role of the inverse Planck
constant. We show this  correspondence by relating the Wigner
distributions of the $N$ fermion systems representing such states,
to low energy space profiles of systems of VSFT D-branes. In this
context the Pauli exclusion principle appears as a consequence of
the VSFT projector equation. The family of 1/2--BPS states
maps through coarse--graining to droplet LLM supergravity
solutions. We discuss the possible meaning of the corresponding
coarse graining in the VSFT side.


\vspace{0.2in} \noindent Keywords: 1/2--BPS states, String Field Theory,
Superstars, Coarse--graining \vspace{0.2in}
\newpage
\section{Introduction}

The essence of this paper is the observation that there is a remarkable
correspondence between states
one meets in the framework of the AdS/CFT duality, and solutions of
vacuum string field theory (VSFT). The correspondence is simply sketched and
is far
from exhaustive (mainly because we do not know enough about supersymmetric
VSFT), but it is very suggestive and, if confirmed, it could lead to very
interesting consequences. Roughly speaking it goes as follows. In the framework
of type IIB superstring theory,
AdS/CFT duality establishes a correspondence between ${\cal N}=4$
superconformal $U(N)$ gauge
field theory on the boundary of $AdS_5$ and supergravity on the background
$AdS_5\times S^5$. In the strongest formulation the correspondence is between
the two theories as a whole. Here we will limit our consideration to a class
of 1/2--BPS states which can be formulated as composite of the $U(N)$ gauge
theory scalars. In general they can be cast in the form of Schur polynomials,
and thus they are in one to one correspondence with
Young diagrams, represented (in the case of giant gravitons \cite{MST}, for instance) by
columns of maximal size (number of boxes) $N$. On the supergravity side
they correspond to 1/2 BPS states that are solutions to the supergravity
equations of motion. The latter represent localized states in the AdS geometry
that wrap around $S^3$ cycles of $S^5$; they are stabilized by their angular
momentum $J$ in $S^5$, with the condition $J\leq N$. There are other
significant 1/2--BPS states with mass $\sim N^2$, the superstars \cite{MyTaf}. 
They correspond
to Young diagrams represented by approximate triangles. On the supergravity side
these are singular 1/2--BPS states with a naked singularity, which are regarded
as solutions on the verge of developing a black-hole horizon due to the quantum
corrections. Giant gravitons and superstars are two examples of
a zoo of new entities that can be constructed in similar ways.

An interesting question is the following one: what is the precise relation
between a state in gauge field theory and the corresponding
supergravity solution? More precisely, how does the geometry that
characterize the latter arise from the former (which, at first sight, is a
totally ungeometrical object)? The answer seems to be coarse-graining:
geometry arises from averaging details of the quantum states in the
gauge theory side.

The argument brought forth by \cite{Bere,BBJS,Silva}
goes as follows.
One remarkable aspect of the above gauge field theory states is that they
can be represented
also in terms of $N$ fermionic oscillators in a harmonic potential. The
correspondence can be once again established via Young diagrams: quantum systems
with the same Young diagram describe the same quantum state. This lends itself
to a very interesting development: to quantum systems of this type we can
associate in a one--to--one way Wigner distribution functions. In general,
to a point--like
system in a $(q,p)$ phase space we can associate a Wigner distribution
function $W(q,p)$. This is nothing but the bosonization of the original fermion
system, but $W(q,p)$ is also very close to a probability
distribution in phase space. In this way we
can associate a Wigner distribution to any state, such as the vacuum,
"black rings", superstars, etc. Now, it so happens that these Wigner
distributions are characterized in the large $N$ limit by (coarse--grained)
profiles that can be matched to the corresponding geometry (droplets) of
the 1/2 BPS supergravity solutions.

With the above premise, the point we want to make in this paper is that
the Wigner distributions
for the above introduced states naturally appear in VSFT. More
precisely, the same profiles appear as low energy space profiles of VSFT
solutions.

Vacuum string field theory (VSFT) is a version of Witten's open string field
theory which is conjectured to represent string theory at the tachyon
condensation vacuum \cite{Ras}. Its action is formally the same as the original Witten
theory except that the BRST charge takes a simplified form: it has been argued
that it can be expressed simply in terms of the ghost creation and annihilation
operators. By virtue of this simplification it has been possible to
determine exact classical solutions which have been shown to represent
D--branes.
The existence of such solutions confirms the conjecture at the basis of
VSFT\footnote{Recently there has been a very important breakthrough in OSFT:
an analytic solution representing the tachyon condensation vacuum has been
determined by Martin Schnabl, \cite{Schn2}. This will certainly
shed new light on tachyon condensation and will hopefully
clarify the role and status (is it only an effective theory?) of VSFT in it.}.
It is possible to introduce families of such solutions, spanned
by Laguerre polynomials. Any sum of distinct solutions is also a solution.
It is also possible to introduce a constant background B--field
in the internal directions and obtain, in the low energy limit, a space profile
for these kind of solutions. The space profiles obtained in this way for a
large family of VSFT solutions are (remarkably) the same as the Wigner
distributions of a corresponding family of half--BPS states introduced above.

This correspondence leads us to a related subject: open--closed string duality
as seen from the SFT point of view. A.Sen has recently conjectured,
\cite{Senopenclosed}, that open string theory might be able
to describe all the closed string physics, at least in a background where
D--branes are present. In this sense VSFT should
be a privileged vantage point: the tachyon condensation vacuum physics can only
represent closed string theory and thus VSFT should be able to describe  closed
string theory in the sense of \cite{Senopenclosed}. The existence of the
D--brane solutions mentioned above is a confirmation of this. However these
D--brane solutions are expressed as squeezed (or related) states. At most,
in the presence of a background B field, we can produce a space profile thereof.
However it has not been known so far how to associate a corresponding geometry.
The correspondence between space profiles and Wigner distributions may be the
clue: by interpreting a space profile as a Wigner distribution, we can
reconstruct a half--BPS state and as a consequence arrive at some definite
geometry, which is the coarse--grained averaging over the corresponding fermion
systems.

The above is the motivation of our research. However it must be said immediately
that we are still far from a complete understanding of all the aspects involved.
The big missing block is supersymmetric vacuum string field theory and its
solutions. Should we know them we would be able to confirm the above suggestions,
which therefore, at this stage, remain mostly conjectural, although not
unmotivated.

The paper is organized as follows. In the next section we give a brief
introduction of 1/2--BPS states in CFT. We insist in particular in their
representation by means of fermion systems of harmonic oscillators. Then we
introduce the corresponding 1/2--BPS supergravity solutions. Finally we
describe in some detail the Wigner distributions, which are conjectured to be,
in the large $N$ limit, the bridge between them. In section 3 we recall what is
relevant in this context about VSFT solutions. In section 4 we establish the
correspondence between $N$ fermion systems and VSFT solutions and make a list
of facts supporting it. Section 5 is devoted to a critical assessment of the
conjectural aspects of our paper.

\section{Half--BPS solutions}

In the field theory side of the AdS/CFT correspondence, half--BPS
multiplets of ${\cal N}=4$ Yang--Mills theory fall into representations
$(0,l,0)$ of the
$SO(6)$ R--symmetry group. Highest weight states can be constructed as gauge
invariant polynomials of a complex scalar field $X$. The conformal dimension of
the latter is $\Delta=1$ and the $U(1)$ R--charge $J=1$, where $U(1)\in SO(6)$.
A highest weight therefore satisfies $\Delta=J$. Basically such states
are constructed out of multiple traces of $X$. The most
general state of this type of charge $n$ takes the form
\be
(\tr(X^{l_1}))^{k_1}(\tr(X^{l_2}))^{k_2}\ldots
(\tr(X^{l_p}))^{k_p}\label{multitrace}
\ee
where the integers $l_i,k_i$ form a partition of $n$: $\sum_{i=1}^pl_ik_i=n$.
A basis for these states is given by the  degree $n$ Schur polynomials of
the group $U(N)$. These in turn correspond to Young tableaux of maximal column
length $N$. Therefore we
can classify these highest weight states (chiral primaries) by means of
Young diagrams \cite{Corley,Aharony,Bere2}. 

It can be shown that they can be represented in another useful way.
In fact their correlators can be related, by using the canonical approach,
by correlators of a suitable (time--dependent) matrix model with a quadratic
potential \cite{Corley,Donos}.
The matrix model can be solved also in another way, by diagonalizing it
and producing in this way a Vandermonde determinant Jacobian factor 
\cite{Bere2}. The latter can be lifted
to the exponential giving rise to a repulsive potential among the eigenvalues.
The result is that we can interpret the eigenvalues $\l_i$ as a system
of $N$ fermionic oscillators with Hamiltonian
$H=\sum_i\l_i^\dagger\l_i+\frac 12$. The energy levels of this system are
given by $E_i=n_i +\frac 12$, where $n_i$ are nonnegative integers.
The corresponding wavefunctions are given in terms of Slater determinants
\be
\Psi(\l_1,\l_2,\ldots,\l_n)\sim\,
e^{-\sum_i \frac{{\lambda_i}^2} 2 }\, \Det
\left(\matrix{H^{n_1}(\l_1)&H^{n_1}(\l_2)&\ldots & H^{n_1}(\l_N)\cr
H^{n_2}(\l_1)&H^{n_2}(\l_2)&\ldots & H^{n_2}(\l_N)\cr
:&:&:&:\cr
:&:&:&:\cr
H^{n_N}(\l_1)&H^{n_N}(\l_2)&\ldots & H^{n_N}(\l_N) \cr}\right)\label{fermwavef}
\ee
where $H_n$ are the Hermite polynomials for a single harmonic oscillator.
The ground state $\Psi_0$ corresponds to $n_1=0,n_2=1,\ldots n_N=N-1$.
Therefore the generic excited state can be represented by means of a
Young diagram with rows $(r_1,r_2,\ldots, r_N)$, with $r_i=n_i-i+1$
not all vanishing natural numbers in decreasing order.
The energy of the state above the Fermi sea is $E=J=\sum_i \, r_i$,
which is the total number of boxes in the Young diagram (for the
relation to Fermi systems, see \cite{LLM,Bere2,Suryanarayana,mandal,Takayama,
Caldarelli}).

Let us list a few states which will be considered in the sequel by means of
their Young diagram representation. A giant graviton is represented by a
single column Young diagram, whose maximum length is of course $N$. A giant
graviton, \cite{MST,Hashimoto,Balasu} is a half--BPS state which can 
be described as a D3--brane wrapping
around an $S^3$ cycles in the $S^5$ factor of $AdS_5\times S^5$. Stability is
guaranteed by the spinning of the brane around an axis in $S^5$. The angular
momentum has an upper bound $J\leq N$, which is a manifestation of the
stringy exclusion principle. Since $\Delta=J$, the representation by means of
a Young tableau incorporates in a simple way the exclusion principle.

A dual giant graviton, i.e.
a D3--brane wrapping around an $S^3$ cycle in $AdS_5$, is represented by
a one--row Young diagram of arbitrary length (no bound here). A black ring
is represented by a large rectangular diagram of size $N$ (see below).
A superstar is represented by a large triangular diagram of size $\sim N$.
It represents a stack of giant gravitons located at the origin of $AdS_5$.
From the supergravity viewpoint, it is a singular solution in that it has
a naked singularity. It is conjectured that due to string corrections
it may actually be completely regular solution.

In the last
two cases the energy of the states is proportional to the area of the Young
tableau and therefore
$\sim N^2$. Following in particular \cite{BBJS}, these are the states we will
be mostly interested in in the following and we will consider them in the large
$N$ limit. In \cite{BBJS} a limit shape was
introduced for the corresponding Young tableaux in the continuous limit.
This is a function $y(x)$, where $x$ runs from left to right along the rows
and $y$ from bottom to top along the columns. The origin is set at the leftmost
bottom box of the tableau. For instance, for the superstar ensemble we have
$\Delta = N N_c/2$ and $y(x) = \frac {N_c}{N} x$, where $N_c$ is the number of
columns.

\subsection{1/2--BPS states as supergravity solutions}

In \cite{LLM} a beautiful characterization of 1/2--BPS states in type IIB
supergravity was found. Regular 1/2--BPS solutions with a geometry invariant
under $SO(4)\times SO(4)\times R$ correspond to the following ansatz
\be
ds^2\!&=&\! -h^{-2} (dt+V_i \,dx^i)^2 + h^2(dy^2 + dx^i\,dx^i) + y\,
e^G d\Omega_3^2 +
y\, e^{-G}d\tilde\Omega_3^2\0\\
h^{-2} \!&=&\! 2y\, cosh\,G\0\\
y\, \d_y V_i\!&=&\!\epsilon_{ij} \,\d_j z,\quad\quad
 y(\d_iV_j-\d_jV_i)=\epsilon_{ij}\,  \d_y z\0\\
 z\!&=&\! \frac 12 tanh\, G\label{halfbps}
 \ee
 where $i,j=1,2$ and $\epsilon_{ij}$ is the antisymmetric symbol.
There are also $N$ units of 5--form flux, with
\be
F_{(5)} = F_{\mu\nu} dx^\mu \wedge dx^\nu \wedge d\Omega_3 +
\tilde F_{\mu\nu} dx^\mu \wedge dx^\nu \wedge d\tilde\Omega_3\0
\ee
where $\mu,\nu=0,...,3$ refer to $t,x^1,x^2,y$. As for the ansatz for $F$
and $\tilde F$, see \cite{LLM}.
The full solution is determined in terms of a single function $z$, which
must satisfy the equation
\be
\d_i\d_i \,z + y\d_y\left(\frac {\d_y z}y \right)=0 \label{zyeq}
\ee
One can solve this equation by remarking its analogy with the Laplace equation
for an electrostatic potential. Regular solutions
can exist only if at the boundary $y=0$ the function $z(0,x_1,x_2)$ takes the
values $\pm \frac 12$. Therefore regular solutions correspond to boundary
functions $z(0,x_1,x_2)$ that are locally constant in the $x_1,x_2$ plane.
The region of this plane where $z= -1/2$ are called droplets and denoted by
${\cal D}$. Following \cite{BBJS} we reintroduce in the game $\hbar$ and
make the identification $\hbar \leftrightarrow 2 \pi \ell_p^4$, noticing that
$x_1,x_2$ have the unusual dimension of a length square.

The area of the droplet must equal $N$:
\be
N= \int_{\cal D} \frac {d^2x}{2 \pi\hbar}\label{qarea}
\ee
while the conformal dimension of the state corresponding to the droplet ${\cal D}$
is
\be
\Delta= \int_{\cal D}\frac {d^2x}{2 \pi\hbar}\,\frac 12 \,\frac {x_1^2+x_2^2}\hbar
- \frac 12\, \left(\int_{\cal D}\frac {d^2x}{2 \pi\hbar}\right)^2\label{Delta}
\ee
In conclusion, the information about the solution is encoded in the droplet.
For instance, if the droplet is a disk of radius $r_0$ we recover the
$AdS_5\times S^5$ solution; if the droplet is the upper half plane one gets the
plane wave solution. In general if the droplet is compact the solution is
asymptotically $AdS_5\times S^5$. It is useful to introduce the new notation
$u(0;x_1,x_2) = \frac 12 - z(0;x_1,x_2)$; $u$ is the characteristic function
of the droplet, since it equals 1 inside the droplet and 0 outside. Solutions
with such (sharp) characteristic functions are regular since the boundary
conditions are satisfied. Solutions characterized by a function $u$ which is
not exactly 1 or 0, are singular \cite{CKS,ML} (for a connection
with quantum Hall effect, see \cite{Bere3,Ghodsi,Boyarski}). This is the case of 
the superstar solution \cite{MyTaf}.

\subsection{The Wigner distribution}

It is clearly of upmost importance to establish a dictionary between the
1/2--BPS states introduced at the beginning of this section starting from
$N=4$ SYM and the droplet solutions. This is tantamount to finding a recipe
to recognize the
geometry emerging from a given gauge field theory state. The clue is the free
fermion representation introduced above: any state represented by a Young diagram
can be interpreted as a system of N fermions with energies above the Fermi
 sea. To  this end it is useful to rewrite the formulas (\ref{qarea}) and
(\ref{Delta}) in the more general form
\be
\Delta \!&=&\! \int\frac {d^2x}{2 \pi\hbar}\,\frac 12 \,
\frac {x_1^2+x_2^2}\hbar
u(0;x_1,x_2)- \frac 12\, \left(\int\frac {d^2x}{2 \pi\hbar}u(0;x_1,x_2)\right)^2
\label{Delta'}\\
N\!&=&\! \int \frac {d^2x}{2 \pi\hbar}u(0;x_1,x_2).\label{qarea'}
\ee
where the integration extends over the whole $x_1,x_2$ plane.
These formulas suggest that $u$ be identified with the semiclassical limit
of the quantum one-particle $(q,p)$ phase--space distributions of the free dual
fermions after the identification $(x_1,x_2)\leftrightarrow (q,p)$.
A phase--space distribution may be understood as an attempt of assigning a
probability distribution to the phase--space points. It is a heuristic concept
and there is no unique prescription for it. The most well--known distribution
is the Wigner one \cite{wigner}:
\be
W(q,p)= \frac 1{2\pi \hbar}\int_{-\infty}^{\infty} dy \langle q-y|\hat \rho
|q+y\rangle
e^{2ipy/\hbar}\label{wigner}
\ee
where $\hat \rho$ is the density matrix. In the case of a pure state $\psi$,
$\langle q'|\hat\rho|q''\rangle= \psi(q')\,\psi^\star(q'')$, therefore
\be
W(q,p)= \frac 1{2\pi \hbar}\int_{-\infty}^{\infty} dy \psi^\star (q+y) \psi(q-y)
e^{2ipy/\hbar}\label{wignerpure}
\ee
In general $\hat \rho$ will take the form of
\be
\hat\rho(q',q'') = \sum_{f\in {\cal F}}\psi_f(q') \psi^\star_f(q'')\0
\ee
${\cal F}$ being a given family of
pure states. We will consider family of pure states representing excited
states of $N$ (fermionic) harmonic oscillators
$f_n= r_n+n-1$, with $n=1,...,N$ (where we have dropped $\hbar$).
In this case ${\cal F}$ will be a subset of the natural numbers and
\be
\psi_{f_n} = A(f_n) H_{f_n}(q/\sqrt{\hbar}) e^{-q^2/2\hbar}\0
\ee
where $A(n)$ is a normalization constant and $H_n$ are, as above,
the Hermite polynomials. Using a well--known integration formula for
Hermite polynomials one gets, \cite{wigner},
\be
W(q,p) = \sum_{f_n\in {\cal F}} W_{f_n}(q,p)=\frac 1{2\pi \hbar}
 e^{-(q^2+p^2)/\hbar}\sum_{f_n\in {\cal F}} (-1)^{f_n}
L_{f_n}\left( 2 \frac {q^2+p^2}\hbar \right)\label{wignerosc}
\ee
Thinking of $W$ as a probability
distribution is certainly a heuristic and approximate concept, because it
may be negative (see the considerations in \cite{BBJS}, where an improved
always positive distribution is introduced, the Husimi distribution).
However we will not need it in the following, because we will compare
Wigner distributions with space profiles of VSFT (which are not probability
distributions either).

Here we are interested in Wigner
distributions because they represent a precise recipe to bosonize
associated fermion systems: from the fermion system we easily get the Wigner distribution and from
the latter we can reconstruct the former, \cite{mandal,dhar}.
In the following we will use Wigner distributions in this sense,
and will be concerned specifically with distributions
relative to ensembles, in which $N$ is supposed to be very large.
The semiclassical limit will correspond to $\hbar \to 0$ keeping $\hbar N$
finite. We will use such distributions to make a comparison with
the $u$ droplet distributions, \cite{BBJS}, and with space profiles in VSFT
(for coarse--graining, see also \cite{Buchel,Shepard,Giombi}).

Let us consider a few significant cases. The first concerns the Fermi sea.
The relevant distribution is
\be
2\pi \hbar W_{FS} = 2\pi \hbar\sum_{n=0}^{N-1}W_n(q,p)\label{FS}
\ee
By using a well--known identity for Laguerre polynomials one formally
obtains 1 when the summation extends to infinity with fixed $\hbar$.
This would not correspond to $AdS_5\times S^5$. However a numerical analysis
shows that the limit $\hbar\to 0$ with $\hbar N$ fixed
reproduces the finite disk characteristic of the latter solution (see, for
instance, \cite{Berry}).

The second example involves the Young diagram corresponding to a giant graviton.
It has $r_n=0, \, n< k$ and $r_n=1$ for $k\leq n\leq N$. The distribution is
\be
2\pi \hbar W_{GG} = 2\pi \hbar\left(\sum_{n=0}^{k-2} + \sum_{n=k}^{N}
\right)W_n(q,p)\label{GG}
\ee
It is evident that in the large $N$ limit with $k$ fixed, this
distribution will be indistinguishable from the Fermi sea one.

The third example is the case corresponding to a rectangular Young diagram
of row length $K$.
It represents $N$ fermions all excited above the sea by the same amount $K$.
There is no a priori relation between $N$ and $K$, but
we are interested in the limit of large $N$ and $K$ such that $\hbar K$
as well as $\hbar N$ are finite. The Wigner distribution is
\be
2 \pi \hbar W_{\rm rect}= 2 \pi \hbar \sum_{n=K}^{N+K-1} W_{K+n-1}(q,p)\label{rect}
\ee
Setting $u(0,x_1,x_2) =2 \pi \hbar W_{\rm rect}$ this identifies a characteristic
function which is (approximately) 1 in the ring $\hbar K\leq \frac {q^2+p^2}2\leq
\hbar (N+K)$ and 0 outside, in the large $N$ and $K$ limit. This corresponds to
the 1/2--BPS called "black ring" in \cite{LLM}. It has conformal dimension
$\Delta=NK\sim N^2$, since $K$ must be some rational multiple of $N$.

The last example concerns Young diagrams which are approximately
triangular with $\Delta = NN_c/2$ and so correspond to superstar ensembles.
In this case we have $f_n= (n-1)\delta_n$, with $\delta_n$ an integer $\sim \frac {N_c}N+1$.
For illustrative purposes let us set $\delta_n=\delta= \frac {N_c}N+1$.
Then
\be
2 \pi \hbar W_{\rm triangle}= 2 \pi \hbar \sum_{n=0}^{N-1}
W_{n\delta}(q,p)
= 2 e^{-\frac{2H}\hbar} \sum_{n=0}(-1)^{n \delta}L_{n\delta}(4H/\hbar)
\label{Wigsuperstar}
\ee
where $H=(q^2+p^2)/2$.
The result of the analysis in \cite{BBJS} is that in the large $N$ limit
\be
2 \pi \hbar W^\infty_{\rm triangle}= \frac 1\delta + {\rm oscillations\, at\,
scale\,\Delta H= \hbar}\0
\ee
Therefore identifying once again $2 \pi \hbar W^\infty_{\rm triangle}$
with $u(0;x_1,x_2)$ we get approximately $u(0,x_1,x_2)= 1/\delta$ within a
finite radius disk. This corresponds to a fractionally filled droplet
and represents the superstar solution,
which is singular. It is also an explicit example of the relations
\be
u(0;r^2)=\frac 1{1+y'}=g(E)\label{grayscale}
\ee
which was conjectured and verified in various examples in \cite{BBJS}.
The function $g(E)$ is called {\it grayscale distribution} and encodes the
effective behavior of coarse--grained semiclassical observables in a given
quantum state.

\section{VSFT: a reminder}

In this section we recall what is strictly necessary from
VSFT in order to render this paper self--contained.
The VSFT action is
\beq
{\cal S}(\Psi)= -  \left(\frac 12 \langle\Psi
|{\cal Q}|\Psi\rangle +
\frac 13 \langle\Psi |\Psi *\Psi\rangle\right)\label{sftaction}
\eeq
where
\beq
{\cal {Q}} =  c_0 + \sum_{n>0} \,(-1)^n \,(c_{2n}+ c_{-2n})\label{calQ}
\eeq
The ansatz for nonperturbative solutions is in the factorized form
\beq
\Psi= \Psi_m \otimes \Psi_g\label{ans}
\eeq
where $\Psi_g$ and $\Psi_m$ depend purely on ghost and matter
degrees of freedom, respectively. The equation of motion splits into
\be
{\cal Q} \Psi_g & = & - \Psi_g *_g \Psi_g\label{EOMg}\\
\Psi_m & = & \Psi_m *_m \Psi_m\label{EOMm}
\ee
where $*_g$ and $*_m$ refer to the star product involving only the ghost
and matter part.\\
The action for this type of solution is
\beq
{\cal S}(\Psi)= - \frac 1{6 } \langle \Psi_g |{\cal Q}|\Psi_g\rangle
\langle \Psi_m |\Psi_m\rangle \label{actionsliver}
\eeq
$\langle \Psi_m |\Psi_m\rangle$ is the ordinary inner product.
We will first concentrate on the matter part, eq.(\ref{EOMm}), assuming the
existence of a universal ghost solution. The solutions are projectors of the $*_m$ algebra.
The $*_m$ product is defined as follows
\beq
_{123}\!\langle V_3|\Psi_1\rangle_1 |\Psi_2\rangle_2 =_3\!\langle
\Psi_1*_m\Psi_2|,
\label{starm}
\eeq
see \cite{GJ1,Ohta, tope, leclair1} for the definition of the three
string vertex $_{123}\!\langle V_3|$. The basic ingredient in this definition
are the matrices of vertex coefficients $V_{nm}^{rs}$, $r,s=1,2,3,\,\,
n,m=1,\ldots,\infty$.

Now we look for solutions that mimic the behavior of the half-BPS
states discussed in the previous section. As it turns out they must
be superpositions of matter projectors (stacks of VSFT D--branes).
The latter have the following characteristics: they must cover
the ordinary 4D Minkowski space (parallel directions), be, in the low energy
limit ($\a'\to 0$), delta--function like
in 16 directions and have some width in the remaining 6 directions (these
22 directions will be referred to as the transversal ones).
Out of the latter two will have a special status, in
that a constant $B$ field will be switched on along them. We can imagine that
all the internal dimensions are compactified on tori, but this is not
strictly necessary for our argument. In the remaining part of
this section we deal with the construction of the solutions and
postpone until the next section a discussion of their connection with
the previous section.

In the following we need both translationally
invariant (in the parallel directions) and non-translationally invariant
solutions (in the transverse directions).

Although there is a
great variety of such solutions we will stick to those introduced in \cite{RSZ2},
i.e. the sliver and the lump. The former is translationally invariant and
is defined by
\beq
|\Xi\rangle = \N e^{-\frac 12 a^\dagger\cdot S\cdot a^\dagger}|0\rangle,
\quad\quad
a^\dagger\cdot S \cdot a^\dagger = \sum_{n,m=1}^\infty a_n^{\mu\dagger} S_{nm}
 a_m^{\nu\dagger}\eta_{\mu\nu}\label{Xi}
\eeq
where $S= CT$ and
\beq
T= \frac 1{2X} (1+X-\sqrt{(1+3X)(1-X)})\label{sliver}
\eeq
with $X=CV^{11}$, where $C_{nm}=(-1)^n\delta_{nm}$ is the so--called twist
matrix.
The normalization constant $\N$ needs being regularized and is formally
vanishing. It has been showed in other papers how this problem could be
dealt with, \cite{BMP1, BMP2}. Our basic projector will have the
form of the sliver along the the space--time directions

The lump solution was engineered to represent a lower dimensional brane
(Dk-brane), therefore it will have ($25-k$)
transverse space directions along which translational
invariance is broken.
Accordingly we split the three string vertex into the tensor product of
the perpendicular part and the parallel part
\be
|V_3\rangle = |V_{3,\perp}\rangle \, \otimes\,|V_{3,_\|}\rangle\label{split}
\ee
The parallel part is the same as in the sliver case while the
perpendicular part is modified as follows.
Following \cite{RSZ2}, we denote by $x^{\bar \mu},p^{\bar \mu}$,
${\bar \mu}=1,...,k$ the
coordinates and momenta in the transverse directions and introduce the
canonical zero modes oscillators
\be
a_0^{(r){\bar \mu}} = \frac 12 \sqrt b \hat p^{(r){\bar \mu}}
- i\frac {1}{\sqrt b} \hat x^{(r){\bar \mu}},
\quad\quad
a_0^{(r){\bar \mu}\dagger} = \frac 12 \sqrt b \hat p^{(r){\bar \mu}} +
i\frac {1}{\sqrt b}\hat x^{(r){\bar \mu}}, \label{osc}
\ee
where $b$ is a free parameter.
Denoting by $|\Omega_{b}\rangle$ the oscillator vacuum
(\,$a_0^{\bar \mu}|\Omega_{b}\rangle=0$\,),
in this new basis the three string vertex is given by
\be
|V_{3,\perp}\rangle'= K\, e^{-E'}|\Omega_b\rangle\label{V3'}
\ee
$K$ being a suitable constant and
\be
E'= \frac 12 \sum_{r,s=1}^3 \sum_{M,N\geq 0} a_M^{(r){\bar \mu}\dagger}
V_{MN}^{'rs} a_N^{(s){\bar \nu}\dagger}\eta_{{\bar \mu}{\bar \nu}}\label{E'}
\ee
where $M,N$ denote the couple of indices $\{0,m\}$ and $\{0,n\}$,
respectively.
The coefficients $V_{MN}^{'rs}$ are given in Appendix B of \cite{RSZ2}.
The new Neumann coefficients matrices $V^{'rs}$ satisfy the same relations as
the $V^{rs}$ ones. In particular one can introduce the matrices $X^{'rs}=
C V^{'rs}$, where $C_{NM}=(-1)^N\, \delta_{NM}$.
The lump solution $|\Xi'_k\rangle$ has the form (\ref{Xi}) with $S$ along
the parallel directions, while $|0\rangle$ is replaced by $|\Omega_b\rangle$
and  $S$ is replaced by $S'$ along the perpendicular ones.
Here $S'=CT'$ and $T'$ has the same form as $T$ eq.(\ref{sliver}) with
$X$ replaced by $X'$. The normalization constant $\N'$ is defined in a way
analogous to $\N$. It can be verified that
the ratio of tensions for such solutions is the appropriate one
for $Dk$--branes. For our basic projector we will choose $k=22$.

As said above, two of the transverse directions are special, in that
a constant background $B$ field is switched on there.
We denote these two directions by the labels $\a$ and $\b$ (for instance
$\a,\b= 24, 25$) and denote them simply by $y_1,y_2$; we take for $B$ the
explicit form
\be
 B_{\a\b}= \left(\matrix {0&B\cr -B&0\cr}\right)
\label{B}
\ee
Then, as is well--known \cite{SW,sugino,KT}, in these two directions
we have a new effective metric $G_{\alpha\beta}$,
the open string metric, as well as
an effective antisymmetric parameter $\theta_{\alpha\beta}$, given by
\be
G_{\a\b} =\sqrt{ {\rm Det} G}\delta_{\a\b},\quad\quad \theta^{\a\b}
=-\epsilon^{\a\b} \theta\label{Gtheta}
\ee
where until further notice we set $\a'=1$ and
${\rm Det} G = \left( 1+ (2 \pi B)^2\right)^2$. In the low energy limit
$\theta \sim 1/B$.
In (\ref{Gtheta}) $\epsilon^{\alpha\beta}$ represents the $2\times 2$
antisymmetric
symbol with $\epsilon^1{}_2=1$. The transverse vertex $|V_{3,\perp}\rangle $
will become in this case $|V'_{3,\perp}\rangle $
\be
|V'_{3,\perp}\rangle = |V_{3,\perp, \theta}\rangle \, \otimes\,|V_{3,\perp}\rangle
\label{split2}
\ee
where
\be
|V_{3,\perp,\theta}\rangle= K_\theta\, e^{-E_\theta}|\Omega_b\rangle\label{V3''}
\ee
$K_\theta$ is a suitable constant and, \cite{BMS1,tope},
\be
E_\theta= \frac 12 \sum_{r,s=1}^3 \sum_{M,N\geq 0} a_M^{(r)\a\dagger}
\EV_{\a\b,MN}^{rs} a_N^{(s)\b\dagger}\label{E''}
\ee
The coefficients $\EV_{MN}^{\a\b,rs}$ are given in \cite{BMS1}.
The new Neumann coefficients matrices $\EV^{rs}$ satisfy the same relations as
the $V^{rs}$ ones. One introduces the matrices $\X^{rs}= C \EV^{rs}$.
Then the lump solution $|\ES\rangle$ along $\a$ and $\b$ has the form (\ref{Xi})
with $|0\rangle$ replaced by $|\Omega_b\rangle$ and $S$ replaced by $\ES$,
where $\ES=C\T$ and $\T$ has the same form as $T$ in eq.(\ref{sliver}) with
$X$ replaced by $\X$. It can be verified that
the ratio of tensions for such solutions is the appropriate one
for $D$--branes in a magnetic field, \cite{BMS2}.

It is possible to construct a full family of such solutions which are $\ast$-- and
$bpz$--orthonormal. This goes as follows, \cite{BMS3,tope}.
First we introduce two `vectors' $\xi=\{\xi_{N\a}\}$ and $\zeta =
\{\zeta_{N\a}\}$, which are chosen to satisfy the conditions
\be
\rho_1 \xi =0,\quad\quad \rho_2 \xi =\xi, \quad\quad
{\rm and}\quad \rho_1 \zeta =0,\quad\quad \rho_2\zeta=\zeta,
\label{xizeta}
\ee
where $\rho_1,\rho_2$ are the half--string projectors \cite{RSZ3,Moeller}.
Moreover we define the matrix  $\tau$ as $\tau = \{ \tau_\a{}^\b\}=
$\mbox{ \small $\left(\begin{array}{cc}1 & 0 \\ 0
& -1 \end{array}\right)$ }. Next we set
\be
\x = (a^\dagger \tau \xi)\, (a^\dagger C \zeta)=
(a_N^{\a\dagger} \tau_\a{}^\beta \xi_{N\b})
(a_N^{\a\dagger}C_{NM}\zeta_{M\a})   \label{xbf}
\ee
Finally we introduce the Laguerre polynomials $L_n(z)$, of the generic
variable $z$, and define the sequence of states
\be
|\Lambda_n\rangle = (-\kappa)^n L_n\Big(\frac{\x}{\kappa}\Big)
|\ES_{\perp\theta}\rangle\label{Lambdan}
\ee
where, for simplicity, we have written down the tensorial factor
involving the the $y_1,y_2$ directions only and understood the other
directions.
As part of the definition of $|\Lambda_n\rangle$ we require the two
following conditions to be satisfied
\be
\xi^T \tau\frac 1{\I-\T^2}\zeta = 1 ,\quad\quad
\xi^T \tau\frac {\T}{\I-\T^2}\zeta= \kappa
\label{cond}
\ee
where $\kappa$ is a real number. To guarantee
Hermiticity for $|\Lambda_n\rangle$, we require
\be
\zeta = \tau \xi^*.\label{real}
\ee

The states $|\Lambda_n\rangle$ satisfy the remarkable property
\be
&& |\Lambda_n\rangle * |\Lambda_m\rangle = \delta_{n,m}
|\Lambda_n\rangle \label{nstarm}\\
&& \langle \Lambda_n |\Lambda_m\rangle = \delta_{n,m}
 \langle \Lambda_0 |\Lambda_0\rangle \label{nm}
\ee
Therefore each $\Lambda_n$, as well as any combination
of $\Lambda_n$ with unit coefficients, are lump solution.

So far we have set $\alpha'=1$. It is easy to insert back
$\alpha'$. In order to evaluate the low energy profile of
$|\Lambda_n\rangle$ we first contract it with the eigenstate of the
position operators with eigenvalues $y^\a$: $\langle y|\Lambda_n\rangle$,
and then take the limit $\alpha'\to 0$, \cite{MT,BMS2,BMS3}. The leading
term in the $\a'$ expansion turns out to be
\be
\langle y| \Lambda_n\rangle  =\frac 1{\pi}(-1)^n\,L_n\left(\frac{2\rho^2}{\theta}\right)
\,e^{-\frac {\rho^2}{\theta}}|\Xi\rangle + {\cal O}(\sqrt{\a'})\label{LElimit}
\ee
where $\rho^2 = y^\a y^\b \delta_{\a\b}$ and $|\Xi\rangle$ is the sliver solution.

The projectors we need in the following have this $\a'\to 0$ limit
in the $y^\a$ directions; as for the remaining directions,
they have the form of the sliver in the
parallel directions and, finally, they become delta--like functions 
multiplied by
the sliver in the
remaining transverse directions, i.e. they are localized at the origin of the
latter. This can be easily seen by taking the limit $\theta\to 0$
in the case $n=0$ in (\ref{LElimit})\footnote{One could easily construct
projectors that are `fat' also along other transverse directions, but we will
not need them in the sequel.}.

\section{A correspondence}

Looking at eqs.(\ref{FS},\ref{rect},\ref{Wigsuperstar}) of section 2, one
immediately notices that they can be seen (up to an overall normalization
constant) as the low energy limit space profiles of combinations of
the string states $\Lambda_n$ introduced in the previous section,
with unit coefficients. Since combinations of $\Lambda_n$ with unit
coefficients are solutions to the equation of motion of VSFT, we can see
the above Wigner distributions as the low energy profile of VSFT solutions
(up to the common $|\Xi\rangle$ factor). It is therefore tantalizing
to make the following association
\be
{\rm Wigner\, distribution\, for\, an\, {\it N}\, fermion\, system} 
\leftrightarrow {\rm VSFT\, solution}\0
\ee
For this to work we must require the correspondence\footnote{It should be
recalled that on the SUGRA side we have three parameters, $\alpha',g_s$ and $N$.
With the first two one forms the combination $\hbar= 2\pi g_s {\a'}^2$. 
On the VSFT side we have also three parameters $\a',\theta$ and $N$.} 
$\hbar \leftrightarrow
\theta$ and that the coordinates $x_1,x_2$ be identified with $y_1,y_2$.
This is what we suggest and would like to motivate in this section.
The previous correspondence can be read in two directions. First: one can say
that to any 1/2--BPS state to which we can associate a Wigner
distribution of the type (\ref{wigner}), there corresponds a VSFT solution
given by a combination
\be
|W_{\cal F}\rangle = \sum_{f_n\in {\cal F}} |\Lambda_{f_n}\rangle,
\quad\quad 2 \hbar W(q,p)|\Xi\rangle= \langle y|W\rangle\,\label{Wassoc}
\ee
where $(p,q)$ is identified with $(y_1,y_2)$ and the latter are the
eigenvalues of $|y\rangle$.
Vice versa: to any VSFT solution of the type (\ref{Wassoc}) we can associate
a Wigner distribution $W(q,p)$ according to (\ref{wigner}). In this way we can
associate
to $|W\rangle$ a Young tableau and therefore a 1/2--BPS state in the ${\cal
N}=4$ superconformal field theory (before taking the large $N$ limit)
and we can associate a geometry (after taking it\footnote{In the process of
taking the large $N$ limit one smears out many details, so that multiple states are mapped to
the same geometry}). The latter point of view
is probably the most interesting one. It implies that we may be able to
associate a geometry to a given VSFT solution, therefore we are in the
condition to answer some of the questions posed by open--closed string duality.
Here we see how geometry emerges from a VSFT solution which is entirely
expressed in terms of open string creation operators.

In the following we would like to list some arguments in support of our
proposal.

1) With the above association we connect a microstate corresponding to
a geometry, which is a supergravity solution, to a string state which is a
solution of the VSFT equation of motion. The correspondence (\ref{Wassoc}) is
one--to--one\footnote{See on this point the remark at the end of section 5.} 
(before the large $N$ limit).

2) The droplet geometry lives in a $(x_1,x_2)$ plane which lies in the
internal (compactified) dimensions. In the same way the plane $(y_1,y_2)$
stays in the compactified part of the bosonic target space. As pointed out
above, we identify the two planes. One could phrase it by saying that the two
space coordinates $x_1,x_2$, which had been replaced by two phase--space
coordinates $q,p$ in the intermediate argument, have returned to their natural
role via the identification with $y_1,y_2$.

3) The correspondence (\ref{Wassoc}) tells us how the Pauli principle gets
incorporated into a bosonic setting. The numbers $f_n$ in the LHS of
(\ref{Wassoc}) correspond to the fermion energy levels in the original  fermion
system. Therefore, due to the Pauli exclusion principle, each $f_n$ can appear
only once in the family ${\cal F}$. Therefore in the summation each
$|\Lambda_{f_n}\rangle$ appears only once. This guarantees that $|W\rangle$
is a VSFT solution\footnote{If $|\Lambda\rangle$ is some $\ast$--projector,
$n|\Lambda\rangle$ is a $\ast$--projector if and only if $n=0,1$.}. On the other
hand any VSFT solution that can be written in the form
$ \sum_{f_n\in {\cal F}} |\Lambda_{f_n}\rangle$ tells us that the numbers
$f_n\in {\cal F}$
can be interpreted as energy levels of a fermionic harmonic oscillator system,
since each appears only once. This is the way the D--brane solutions of VSFT
manifest their fermionic nature.

4) The VSFT solution corresponding to the Fermi sea (\ref{FS}) is represented
by a stack of $N$ (unstable) VSFT D--branes. The giant graviton solution
(\ref{GG}) is represented
by a D--brane missing from the stack. The superstar solution (\ref{Wigsuperstar}) is
represented by a stack of such missing (unstable) D--branes. This is
in keeping with the interpretation of superstars as stack of giant gravitons,
\cite{MyTaf}. (It is worth remarking at this point that all the VSFT solutions
we consider in this paper are composite of VSFT D--branes and there is no
direct identification between single VSFT D--branes and single D3--branes in
superstring theory.)

5) There is an algebra isomorphism between Wigner distributions of the type
(\ref{wigner}) and VSFT solutions like $|W\rangle$, an isomorphism that
was pointed out in \cite{W2,Sch,BMS3}. It is a well--known fact that any
classical function $f(q,p)$ in a $(q,p)$ phase space can be mapped to a quantum
operator $\hat O_f$ via the Weyl transform, and that the product
for quantum operators $\hat O_f \hat O_g$ is mapped into the Moyal product
$f\star g$ for functions. Under this correspondence the $(x_1,x_2)\leftrightarrow
(q,p)$ plane becomes noncommutative. It is a well--known fact that,
under this correspondence the classical Wigner distributions like
(\ref{FS},\ref{rect},\ref{Wigsuperstar}) are mapped into projectors of the
Moyal star algebra:
\be
(2\pi \hbar W) \star (2\pi\hbar W)=2\pi \hbar W.\label{proj}
\ee
Actually these distributions turn out to
coincide with families of the so--called GMS solitons, \cite{GMS,Komaba}.
Let us recall the relevant construction. Define the harmonic oscillator
$a = (\hat q +i \hat p)/\sqrt{(2\theta)}$ and its hermitian conjugate
$a^\dagger$:
$[a,a^\dagger]=1$. The normalized
harmonic oscillator eigenstates are:
$|n\rangle = \frac {(a^\dagger)^n}{\sqrt n!}|0\rangle$.
Now, via the Weyl correspondence, we can map any rank one projector
$|n\rangle\langle n|$
to a classical function of the coordinates $x_1,x_2$.
\be
|n\rangle\langle n|\longleftrightarrow
\psi_n(x_1,x_2)=  2 \, (-1)^n L_n \bigg(\frac {2r^2}\theta\bigg)
e^{- \frac {r^2}\theta}\label{solitons}
\ee
where $r^2 = x_1^2+x_2^2$. Each of these solutions, by construction, satisfy
$\psi_n\star\psi_n =\psi_n$. These are referred to as GMS solitons \cite{GMS}.
In the previous section we have shown that the low energy limit of
$\langle y|\Lambda_n\rangle$ factorizes into the product of the sliver
state and $\psi_n(y_1,y_2)$. This means that the VSFT star product factorizes
into Witten's star product and
the Moyal $\star$ product, \cite{W2,Sch}. More precisely we can formalize the
following isomorphism
\be
\matrix{|\Lambda_n\rangle &\longleftrightarrow&
P_n
&\longleftrightarrow&\psi_n (y_1,y_2)  \cr
|\Lambda_n\rangle * |\Lambda_{n'}\rangle &\longleftrightarrow&
P_n P_{n'}&\longleftrightarrow &\psi_n \star \psi_{n'}}\label{c2}
\ee
where $\star$ denotes the Moyal product.

This remark should not be underestimated. Let us consider $2\pi\hbar W$
and suppose it is such that we can ignore its derivatives with respect to
$p$ and $q$. Then eq.(\ref{proj}) becomes $(2\pi\hbar W)^2= 2\pi\hbar W$,
which is the equation of a characteristic function (it can only be either
0 or 1).
This is indeed what happens in the case of the vacuum and the black ring
solutions, see \cite{BBJS,Silva}. It is not the case of the superstar
distribution because in that case we cannot ignore derivatives. But this
remark suggests that the property of being Moyal projectors is
basic for Wigner distributions to represent 1/2--BPS states.
The string state $|W\rangle$ `inherits' this property, it is the
`continuation' of the space profile to the whole string theory. In this sense
it is natural that $|W\rangle$ be a string field theory solution.

6) Finally one should point out that there exists a solution generating
technique that allows one to produce new solutions starting from a
fixed one. As an
example let us consider the partial isometries introduced in \cite{carlo}.
They are defined as follows
\be
({\cal P}_+)^k \ket{\Lambda_n}=\ket{\Lambda_{n+k}}\\
({\cal P}_-)^k \ket{\Lambda_n}=\ket{\Lambda_{n-k}}\\
\ee
where we define $\Lambda_n=0$ for $n<0$ (see \cite{carlo} for definitions
of ${\cal P}_\pm)$). Given a state represented by a certain Young diagram,
the operator $({\cal P}_+)^k$ adds k boxes in each of the $N$ rows, while its
inverse $({\cal P}_-)^k$ removes them. Consider for instance the Fermi sea
solution corresponding to eq.(\ref{FS}) and apply to it $({\cal P}_+)^K$.
According the above equations one gets the solution corresponding to the
rectangular Young diagram (\ref{rect}). We have seen that (\ref{rect}), in the
large $N$ limit, leads to the black ring geometry.
Note that the transformations that are
generated by such partial isometries are area preserving on the droplet plane
(the number of fermions is left unchanged). This
points to the fact that partial isometries on the open string side are
mapped to topology changing transformations on the
closed string side.

\subsection{Matching observables}

In this subsection we deal with the identification of the quantities
in VSFT that correspond to two basic observables in the superconformal 
field theory and in the supergravity side. The latter are given
by the total five form flux, $N$, and by the energy, $\Delta$, 
(\ref{qarea}, \ref{Delta}).
We would like to see how these two observables are encoded in
the star algebra that characteres the VSFT solutions.

The total flux is simply given by the bpz norm of the projector
corresponding to the given Young diagram.
We have indeed, (\ref{nstarm}),
\be\label{N}
N=\frac{\bra {W_{\cal F}}W_{\cal F}\rangle}{\bra{\Lambda_0}\Lambda_0\rangle}
\ee
This is perfectly expected as the total flux is determined by the number of 
boundary branes producing it. Note that, differently from the usual open 
string description given by the gauge theory, this observable {\it is not} 
part of the definition of the theory but is part of a classical solution 
(in much the same way as it happens in gravity).

In order to understand how the observable corresponding to $\Delta$ emerges 
from the star algebra, an extension of the (\ref{nstarm}) is necessary. 
Consider the following non twist invariant states, \cite{tope}
\be
|\Lambda_{n,m}\rangle \,&=&\, \sqrt \frac {n!}{m!} (-\kappa)^n \,
Y^{m-n} L_n^{m-n}
\left(\frac \x \kappa\right)|\ES_{\perp\theta}\rangle,
\quad\quad n\leq m \label{nm1}\\
|\Lambda_{n,m}\rangle \,&=&\, \sqrt \frac {m!}{n!} (-\kappa)^m \,
X^{n-m} L_m^{n-m}
\left(\frac \x \kappa\right)|\ES_{\perp\theta}\rangle,
\quad\quad n\geq m\label{nm2}
\ee
where
\be
X= a^\dagger \tau \xi \, \quad\quad Y = a^\dagger C\zeta\label{XY}
\ee
so that $\x = XY$,
and $L_n^{m-n}(z) = \sum_{k=0}^m
\left(\matrix {m\cr n-k}\right) (-z)^k/k!$ are the generalized Laguerre 
polynomials. Note that $\Lambda_n=\Lambda_{nn}$.

These states star--multiply among themselves in the following way

\be
\Lambda_{nm}*\Lambda_{pq}=\delta_{mp}\Lambda_{nq}
\ee
In VSFT they have been shown to be the vacuum states for
perturbative strings stretched between the n-th and the m-th brane, 
\cite{carlo}

Thanks to this extended algebra we can explicitly realize the 
fermionic system of our concern. To this end let's define the 
following {\it inner} operators acting on the string Hilbert space
\be
A_+\phi&=&\sum_{n=0}^{\infty} \sqrt{n+1}\; 
\Lambda_{n+1,n}*\phi*\Lambda_{n,n+1}\0\\
A_-\phi&=&\sum_{n=0}^{\infty} \sqrt{n+1}\; 
\Lambda_{n,n+1}*\phi*\Lambda_{n+1,n}\0
\ee
for any string state $\phi$.
These  are  (the adjoint representation of) the string field oscillators 
defined in \cite{okuhalf} (at {\it fixed} half string vector) and
behave as the raising/lowering operators of a harmonic oscillator
\be
A_+\Lambda_n&=&\sqrt{n+1}\;\Lambda_{n+1}\0\\
A_-\Lambda_n&=&\sqrt n\;\Lambda_{n-1}\0
\ee
It is then natural to consider  the operator
\be
H=A_+\,A_-\label{H}
\ee
which, up to zero point energy, is the analog of the harmonic oscillator 
hamiltonian.

For single brane states we have
\be
H\,\Lambda_n=n\; \Lambda_n\0
\ee
and, more important,
\be
n=\frac{\bra{\Lambda_n}H\ket{\Lambda_n}}{\bra{\Lambda_0}\Lambda_0\rangle}\0
\ee
If we evaluate this operator on the stack $W_{\cal F}$ we get
\be
\frac{\bra {W_{\cal F}}H\ket{W_{\cal F}}}{\bra{\Lambda_0}\Lambda_0\rangle}=
\sum_{n=0}^{N-1}f_n
\ee
This is nothing but the energy of the corresponding fermion ensemble given 
by the Young diagram $\cal F$.

Now let us define the observable that corresponds to $\Delta$, which we
will denote with the same symbol.
As in the gravity side $\Delta$ is defined as the 
difference in `energy' between
the state under consideration and the `vacuum' (empty AdS), 
at {\it fixed} five form flux $N$, see (\ref{Delta}).
\be
\Delta=\frac{\bra {W_{\cal F}}H\ket{W_{\cal F}}-
\bra {W_{{\cal F}_0}}H\ket{W_{{\cal F}_0}}}{\bra{\Lambda_0}\Lambda_0\rangle}
=\texttt{number of boxes of the Young Tableau}\label{Delta''}
\ee
where ${\cal F}_0=\{0,1,...,N-1\}$ is the Fermi sea at given $N$.

We remark that in open string theory $N$ in (\ref{N}) is interpreted 
as an energy, while $\Delta$ in (\ref{Delta''}) is not (but it can be 
intepreted as energy in the closed string side under the open--closed 
string duality, \cite{prep}).

\section{Discussion}

The correspondence (\ref{Wassoc}) in the previous section is based on a
series of facts, which have been listed above. The coincidence might be
accidental, but we tend to believe it has a deeper meaning.
The suggestion that comes from the previous section is summarized in the
following table:
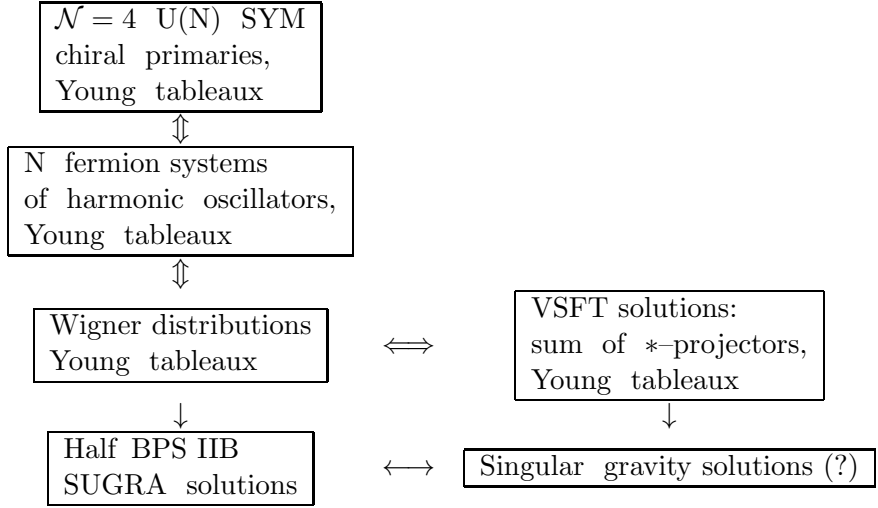
\be
\matrix{
\begin{tabular}{|l|} \hline
${\cal N}=4$\,\, U(N)\,\, {\rm SYM} \\{\rm chiral\,\, primaries,}\\
{\rm Young\,\, tableaux}\\ \hline
\end{tabular}&{}&{} \cr
\Updownarrow &{}&{}\cr
\begin{tabular}{|l|} \hline
 {\rm N \,\,fermion\,\,systems} \\{\rm of\,\, harmonic\,\, oscillators,}\\
 {\rm Young\,\, tableaux}\\  \hline
\end{tabular}&{}&{} \cr
 \Updownarrow &{}&{}\cr
\begin{tabular}{|l|} \hline
 {\rm  Wigner\,\,distributions}\\ {\rm Young\,\, tableaux}\\\hline
\end{tabular}&\Longleftrightarrow&\begin{tabular}{|l|}\hline
 {\rm  VSFT\,\,solutions:}\\ {\rm sum \,\,of \,\,$\ast$--projectors, }\\
 {\rm Young\,\, tableaux}\\ \hline
\end{tabular}\cr
\downarrow &{} &\downarrow \cr
\begin{tabular}{|l|} \hline
 {\rm  Half \,BPS\,\,IIB} \\{\rm SUGRA\,\, solutions}\\ \hline
\end{tabular}&\longleftrightarrow &\begin{tabular}{|l|} \hline
 {\rm  Singular\,\, gravity\,\,solutions (?)}\\ \hline
\end{tabular}
\cr}\0
\ee
where double--line arrows represent one--to--one correspondences, simple
down arrows represent the large $N$ limit and the question mark indicates
the conjectural part of our proposed correspondence. Let us describe it
in more detail.

The fact that $|W\rangle$ is a VSFT solution is the strongest support of
our conjectured correspondence. The weak point  is that
we know it is a solution of bosonic VSFT but we do not know whether it is
a solution of the supersymmetric vacuum string field theory. However we would
like to notice that the 1/2--BPS states considered in \cite{Bere,BBJS,Silva}
in the gauge theory side, are all (very heavy) bosonic states. It is not
unconceivable that the bosonic part of 1/2--BPS states is well described
by solutions of the bosonic string theory. Unfortunately
the study of the tachyon condensation in superstring field theory has not
progressed much, see \cite{aref,ohmori}. From what we know nowadays it is
possible that the bosonic
parts of some solutions of supersymmetric VSFT take a form like $|W\rangle$,
although a complete solutions has not yet been determined.

This raises a problem as to the interpretation of the lower right corner
of the above table. Based on the above argument, they should represent
the bosonic part of supergravity solutions. Now the Einstein
equation for the latter is
\be
R_{\mu\nu}\sim
F_{\mu\l_1\l_2\l_3\l_4} F_{\nu}{}^{\l_1\l_2\l_3\l_4}\label{EOMSUGRA}
\ee
where  $F$ is the five--form field strength and where we have set the
dilaton to 0. The contribution of the RHS is basic
in the case of LLM solutions, as the latter do not satisfy
the pure gravity equation $R_{\mu\nu}=0$. As a consequence
the solutions in the lower right corner above are not pure gravity solutions.
This is the problem we alluded to above: from a purely bosonic theory we get,
in the low energy limit, (the bosonic part of )supergravity solutions.
The most likely explanation
of this surprising result lies in the type of limit we have taken in VSFT:
first $\alpha'\to 0$ with $\theta$ fixed, and then $\theta\to 0$ with
$\theta N$ fixed. This two--step limit selects the (bosonic part of the)
droplet--like supergravity solutions. But, on the VSFT side, there are other
possible limits. For instance, one could take the same limit, but in three
steps, first $\a'\to 0$, then $\theta\to 0$ and finally $N\to \infty$.
This shrinks the droplets to zero size and leads to singular (in the sense
of delta--like) solutions, which we can identify with (singular)
pure gravity solutions. However these solutions are too singular to base on them
any serious discussion. So let us deal with this subject from another viewpoint.

From the above we see that the parameter $\theta$, i.e. the $B$ field,
plays a fundamental role in the VSFT side limit. There is no $B$ field in the
IIB supergravity solutions side, there is instead
a background five--form field, whose flux in suitable units equals $N$.
Of course in the bosonic SFT side there cannot be any such background. However
we have seen that $\theta\sim 1/B$ is identified with $\hbar$ and, in the large
$N$ limit, $\hbar N$ is kept finite. Therefore $B$ and the five--form flux play
a parallel role. The five--form flux supports the supergravity solution,
while $B$ supports the corresponding VSFT solution, which otherwise would
collapse to a delta function. It looks like $B$ is a surrogate of the
five--form flux in the bosonic theory (see the considerations about
noncommutativity in \cite{Das}). One may wonder whether this remark can be
confirmed in some independent way. There actually exists a way in the gravity
side, although it is very hard to verify it analytically.

The low energy equations of motion for bosonic string theory is
\be
R_{\mu\nu}\sim H_{\mu\lambda\rho} H_{\nu}{}^{\lambda \rho}
\label{EOMGRA}
\ee
where again we have set the dilaton to 0, and $H=dB$. Comparing this with
(\ref{EOMSUGRA}) we see that the $H^2$ term might play the role
of the five--form term there. There is however an obstacle. Our
VSFT solutions contain a constant B--field and if we replace a constant
$B$ in (\ref{EOMGRA}) the $H^2$ term vanishes. Nevertheless there exists
another interpretation. Let us consider for definiteness the vacuum
solution, where the droplet is a disk of finite radius $r_0$ in the phase
space. As we remarked above we can easily reproduce this solution on the
VSFT side with a constant field $B=B_0$, by taking the limit $N\to \infty$ such
that $\theta_0 N \sim r_0$. Now, the region $r>r_0$ corresponds to a vanishing
distribution in the large $N$ limit. This can be reproduced as well by taking
first $\theta \to 0$ and {\it then} $ N\to \infty$. In other words, in order
to reproduce the vacuum solution it is not necessary to assume a constant
B field everywhere. Unfortunately we do not know how to deal with VSFT in the
presence of a varying background $B$ field. But we believe it is reasonable to
assume that, anyhow, well inside the droplet the VSFT solution will be
described by the solution with $B=B_0$, far outside the droplet by a solution
with very large $B$ and in the intermediate region by some interpolating
solution. If this is correct then our overall VSFT solution will correspond to
a non constant $B$ field: after coarse--graining the profile will be such that the
$H^2$ term in (\ref{EOMGRA}) mimics the five--form quadratic term in
(\ref{EOMSUGRA}). Of course one should take care also of the
other gravity equations of motion: in order to satisfy them all a nontrivial
dilaton might be necessary and the solution is anyhow very likely to be singular.
One may object that we are here in presence of two different
VSFT solutions (with constant and non constant $B$ field) that correspond to
the same space profile. This is true, but the profile we have been talking is
the same only in the $\a'\to 0$ limit. We expect the $\a'$ corrections to
remove the degeneracy between them.

In any case using bosonic string field theory to establish a correspondence
with 1/2--BPS states of an ${\cal N}=4$ YM theory can be taken, at this stage,
only as a suggestion. However the elements we have listed above are striking.
So let us suppose that our conjecture is correct at least for the class
of states and solutions we are interested in.
Then it is convenient to view it in the framework of
open--closed string duality. One should not forget that VSFT is a version of
{\it open} string field theory, i.e. its language is the language of open
string theory. The correspondence we have established above is between
1/2--BPS states of SYM theory in 4D and {\it full} VSFT solutions. This
suggests that the open string field theory we have been considering describes
in fact the physics of open strings attached to the stack of D3--branes
where the SYM is defined, that is it is the stringy completion of the latter
theory. Once again, the appropriate treatment should
make use of superstring field theory. But let us suppose that bosonic SFT
is enough for the present purpose; then we must conclude that, using
tachyon condensation, we have found a way to pass from open SFT solutions
to space time geometry (via coarse--graining), which is one of the
major problems in open--closed string duality, and this with the additional bonus
of the $\a'$ corrections.

We would like to add two specifications.
The first concerns Chan--Paton factors one is expected to introduce
in order to represent a $U(N)$ theory, and we have not. This is in fact
not necessary, since it was proven in \cite{carlo} that VSFT contains solutions
with all type of $U(N)$ CP factors without the need to introduce them by hand.
The second concerns the string critical dimension, which, in the bosonic case,
is D=26, while the physics of SYM theory lives in D=4. However we have seen
in section 3 that our VSFT solution spontaneously solve the problem, because
we can choose them translationally invariant in 4D and of finite or
zero size in the transverse dimensions, as need be.

In this paper we have only considered Wigner distributions. As pointed out in
\cite{BBJS} there are other proposals, for instance the Husimi distribution,
which is based on a convolution of the Wigner one. On the other hand there
are many other solutions in VSFT, beside the family based on the sliver and
the Laguerre polynomials we have considered so far, for instance the family
of butterfly projectors. It would be interesting to see whether our
correspondence extends to
other phase--space distributions and to other VSFT solutions.

Concerning the future problems to be studied an interesting one relates
to the possible utilization of the full VSFT to calculate $\a'$ corrections.
It has been suggested that the superstar solution may develop a horizon
due to the stringy corrections. Now, a string state like $|W\rangle$ contains
the $\a'$ corrections to its low energy profile. It is therefore natural to
ask whether this knowledge can be translated to the supergravity side. Related
to this is the problem of counting the microstates in order to evaluate the
entropy of the ensemble, \cite{Bere,BBJS,Silva}. To this end one should be able
to count the distinct string fields corresponding to a given low energy
profile. A problem nested into this is related to gauge equivalence. The states
$|\Lambda_n\rangle$ are defined in terms of a vector $\xi_n$
(likewise for the ghost part). These infinitely many
numbers $\xi_n$ are irrelevant in the low energy limit. This fact
is understood in the sense that these numbers  are likely to represent
only gauge degrees of freedom (otherwise also our previous claim of
one--to--one correspondence would seem to need a better phrasing).
It would be interesting to find a real proof of this.

\vskip 1cm

\begin{center}
{\bf Acknowledgments}
\end{center}
L.B. would like to thank A.P.Balachandran, D.Karabali, V.P.Nair and
A.P.Polychronakos for useful remarks.
This research was supported by the Italian MIUR under the program
``Teoria dei Campi, Superstringhe e Gravit\`a'' and by
CAPES-Brasil as far as R.J.S.S. is concerned. The work of CM is
partially supported by IISN - Belgium (convention 4.4505.86), by
the ``Interuniversity Attraction Poles Programme -- Belgian
Science Policy " and by the European Commission programme
MRTN-CT-2004-005104, in which he is associated to V.U. Brussel.


\end{document}